\documentclass[12pt,psfig]{article}
\usepackage{graphicx,epsfig}
\newcommand{\ba}{\begin{array}}
\newcommand{\ea}{\end{array}}
\newcommand{\beq}{\begin{eqnarray*}}
\newcommand{\eeq}{\end{eqnarray*}}
\newcommand{\bdm}{\begin{displaymath}}
\newcommand{\edm}{\end{displaymath}}

\begin{document}
\baselineskip=18pt
\renewcommand {\thefootnote}{\dag}
\renewcommand {\thefootnote}{\ddag}
\renewcommand {\thefootnote}{ }

\pagestyle{plain}

\begin{center}

\vspace{-0.65 in} {\Large \bf A discrete computer network model with expanding dimensions$^*$}\\
[0.3in]

Yuming Shi $^{1}$, Guanrong Chen $^{2\,**}$

\vspace{0.15in}
$^1${\it Department of Mathematics, Shandong University \\[-0.5ex]
Jinan, Shandong 250100, P.~R. China}

\vspace{0.15in} $^2${\it Department of Electronic Engineering\\[-0.5ex]
City University of Hong Kong, Hong Kong SAR, P.~R. China}
\end{center}

\footnote{\ \ Email addresses: ymshi@sdu.edu.cn (Y. Shi),
gchen@ee.cityu.edu.hk (G. Chen).} \footnote{$^{*}$ This research was
supported by the NNSF of Shandong Province (Grant Y2006A15), the
NNSF of China (Grant 10471077), and the Hong Kong Research Council
under the CERG Grant CityU 1114/05E.} \footnote{$^{**}$ The
corresponding author.}\

\vspace{0.15in}

\baselineskip=20pt
\vspace{0.15in}

\noindent{\hspace{0.0in} {\large \bf \hspace{-0.15in} Abstract.}\
Complex networks with expanding dimensions are studied, where the
networks may be directed and weighted, and network nodes are varying
in discrete time in the sense that some new nodes may be added and
some old nodes may be removed from time to time.  A model of such
networks in computer data transmission is discussed. Each node on
the network has fixed dimensionality, while the dimension of the
whole network is defined by the total number of nodes. Based on the
spectacular properties of data transmission on computer networks,
some new concepts of stable and unstable networks differing from the
classical Lyapunov stability are defined. In particular, a special
unstable network model, called devil network, is introduced and
discussed. It is further found that a variety of structures and
connection weights affects the network stability substantially.
Several criteria on stability, instability, and devil network are
established for a rather general class of networks, where some
conditions are actually necessary and sufficient. Mathematically,
this paper makes a first attempt to rigorously formulate a
fundamental issue of modeling discrete linear time-varying systems
with expanding dimensions and study their basic stability property.
\vspace{0.1in}

}
\bigskip

\newpage

\noindent {\large \bf 1.\, Introduction}

\vspace{0.10in}

Many real-world networks appear to be different but share some
similar complexity in such diverse aspects as varying
dimensionality, intrinsic connectivity, and complicated dynamics.
In the endeavor of understanding the common forming mechanism of
seemingly different networks, some recent work has already
captured some essential features of various complex networks,
particularly the small-world characteristic coined by Watts and
Strogatz (1) and the scale-free degree distribution in an
invariant power-law form discovered by Barab\'asi and Albert (2).
It is noticeable that owing to the mathematical simplification for
elegancy and rigor, the classical random graph theory of Erd\"os
and R\`enyi (3) and the small-world model which inherits the same
spirit have a fixed dimensionality for each given network.
Although the scale-free model generates a growing network, the
network dimension is typically restricted to be fixed as that of a
single node in order to be mathematically manageable when analysis
comes into play.

The main interest in this letter is to deal with a growing network
with expanding dimensionality, a more realistic model for the
Internet to say the least. The aim is to build a graph framework and
to lay a mathematical foundation for such a model that evolves in
discrete time with increasing dimensions as well as complex dynamics
including directionality and weights if desirable. A typical example
in point is the real data transmission on a computer network, in
which the number of nodes and the connection among them are both
varied in time, where the connectivity may also be directed and
weighted. Indeed it is quite interesting to think of an email
network in a university or a company, where one computer
contaminated with reinfection-enhanced virus such as the infamous
W32/Sircam will send out many copies of the virus to computers
listed in its email address book, causing the system server
overloaded as an immediate consequence.

A new graph model for such networks as the aforementioned computer
system is established in this letter, where each node on the network
has fixed dimensionality while the dimension of the whole network is
defined by the total number of its nodes therefore continuously
increasing. The network is characterized by a discrete linear
dynamical system, where some new nodes will be added and some old
nodes will be removed throughout the evolutionary process. New
concepts of stable and unstable networks are introduced, which
differ from the classical Lyapunov stability in several aspects. In
particular, a special unstable network model, named devil network,
is discussed. It is shown that a variety of structural and
connectional properties affects the network stability substantially.
Several criteria on stability, instability, and the devil network
are finally established, actually for a rather general class of
networks, where some conditions are necessary and sufficient. The
major mathematical contribution of this paper is to rigorously
formulate a fundamental issue of modeling discrete linear
time-varying systems with expanding dimensions and study their basic
stability theory.

\bigskip

\noindent {\large \bf 2.\, A graph model of simple computer
networks} \vspace{0.10in}

Consider an isolated local-area computer network with only one
server for simplicity, assuming that at most one PC is being added
to the network at a time. In the model, connections among nodes
are directed and the directions may vary in time, but
bi-directional data transmissions are not permitted.

In a real-world network, some new nodes may be added and some old
nodes may be removed from time to time. When a node is removed
from the network at some time $t_0$, one treats it as an isolated
node starting from $t_0$. This means that those removed nodes will
not have any connections with the other nodes in the network for
all $t\ge t_0$, and consequently all the corresponding connection
weights become $0$ forever.

Assume that the computer network has $n_t$ computers, referred to
as nodes, at discrete time $t\in {\mathbf
Z}^+=\{t\}_{t=0}^{\infty}$. Let $x_i(t)$ be the difference in data
amount between the input and output of node $i$, $1\le i\le n_t$,
at time $t$, whose absolute value $|x_i(t)|$ is called the storage
of node $i$ at time $t$. Then
 $$\Delta(x)(t)=\sum_{i=1}^{n_t} x_i(t)$$
is the total difference between the input and output data of the
whole network at time $t$. Note that this $\Delta(x)(t)$ is also
the difference between the output and input data in the server,
referred to as the central station, at time $t$, and its absolute
value $|\Delta(x)(t)|$ is called the storage of the server or the
whole network at time $t$. Only the case of finite memories is
considered; namely, every computer and the server have a maximum
allowable storage.

Conceptually, if the amount of data stored on a computer is less
than or equal to its maximum storage and the amount of data stored
on the server does not exceed its maximum at some time, then the
network is running well at that moment. The network is said to be
stable if it runs well at all times. Otherwise, if the amount of
data stored on some computer is larger than its maximum storage at
some time, whenever in the process, the network is in a
troublesome situation since it would require the server or some
other computers to share the extra workloads. There is another
scenario that the actual storages of all computers are less than
or equal to their maximum allowable storages, but the server is
overloaded, at a moment. In this latter case, the server will
breakdown. Both of these two cases of the network are referred to
as being unstable.

Mathematically, the above concepts of stability and instability
are defined for the model as follows. Let $S_i$ be the maximum
storage of node $i$ and $M_0$ the maximum storage of the server
(or the whole network).

\medskip

\noindent {\bf Definition 1.} A network is said to be stable if
there exists a positive constant $r_0\le M_0$ such that for all
initial point $x(0)=\Big(x_1(0),x_2(0),\ldots,x_{n_0}(0)\Big)^T\in
{\mathbf R}^{n_0}$ satisfying $\sum_{j=1}^{n_0} |x_j(0)| \le r_0$,
one has $|x_i(t)|\le S_i, \;1\le i\le n_t$, and $|\Delta(x)(t)|\le
M_0$, for all $t\in {\mathbf Z}^+$. Otherwise, it is said to be
unstable. In particular, the network is called a devil network if
it is unstable and, further, for any small positive constant $r\le
M_0$ there exists an initial point $x(0)\in{\mathbf R}^{n_0}$ with
$\sum_{j=1}^{n_0} |x_j(0)|\le r$ such that $|x_i(t)|\le S_i,
\;1\le i\le n_t$, for all $t\in {\mathbf Z}^+$, and
$|\Delta(x)(t_k)|\le \alpha M_0$ for infinitely many times
$t_k>0$, $k\ge 1$, where the constant $0<\alpha < 1$ is called a
scaling parameter of the network, but $|\Delta(x) (t_k')|> M_0$
for infinitely many times $t_k'>t_k$, $k\ge 1$.

\medskip

\noindent {\bf Remark 1.}
\begin{itemize}
\item[{\rm (i)}]
The above-introduced definition of stability for networks is
different from the classical Lyapunov stability for dynamical
systems (4).

\item[{\rm (ii)}] In the above definition of the devil network,
the constant $\alpha$, $0<\alpha < 1$, is determined by some
specific requirements on the network. For example, one may choose
$\alpha=1/2$ in the data transmission model of the computer
network. If $|x_i(t)|\le S_i, \;1\le i\le n_t$, for all $t\in
{\mathbf Z}^+$, and $|\Delta(x)(t_k)|\le M_0/2$ for infinitely
many times $t_k$, then each computer runs very well at any time
and the whole network works fine at all $t_k$. But, in the case of
$|\Delta(x)(t_k')|> M_0$ with $t_k'>t_k$, the network would be in
a rapidly changing troublesome situation (devil behaviors) after
$t_k'-t_k$, $k\ge 1$.

\item[{\rm (iii)}] The condition $t_k'>t_k,\;k\ge 1$, in the
definition of devil network, is not restrictive. Since
$\{t_k\}_{k=1}^{\infty}$ and $\{t_k'\}_{k=1}^{\infty}$ are both
infinite sequences, one can easily choose suitable $t'_{k'}$
satisfying $t'_{k'}>t_k$, $k'>k\ge 1$.
\end{itemize}

Define the dimension of the network at time $t$ be equal to $n_t$,
the total number of all the nodes in the network at time $t$.
Ignoring nonlinear factors, simply assume that $x(t)=(x_1(t),x_2(t),
\ldots, x_{n_t}(t))^T\in {\mathbf R}^{n_t}$ satisfies the following
discrete linear system:
 $$x(t+1)=A(t)x(t),\;t\in {\mathbf Z}^+, \eqno{[1]}$$
where $A(t)=(a_{ij}(t))$ is the coupling matrix of the network,
which is an $n_{t+1}\times n_t$ matrix. Given any initial point
$x(0)\in {\mathbf R}^{n_0}$, the solution $x(t)$ of system
${\mathbf 1}$ can be written as
 $$x(t+1)=D(t)x(0),\;\;t\in {\mathbf Z}^+,                \eqno{[2]}$$
where
 $$D(t)=A(t)A(t-1)\cdots A(0).                  \eqno{[3]}$$

Consider the following special case of the above network model in
the rest of this section, where the matrix $A(t)$ has entries
taken from the triple $\{-1,0,1\}$ for all $t\in {\mathbf Z}^+$,
which is called a T-matrix:

\begin{itemize}
\item[{(H$_1$)}] The matrix $A(t)$ of system ${\mathbf 1}$ is an
$n_{t+1}\times n_t$ T-matrix. Assume that its entries $a_{ij}(t),\;
i\ne j$, are evaluated at $t$ in the following way:\ $a_{ij}(t)=-1$
if node $i$ sends data to node $j$; $a_{ij}(t)=1$ if node $i$
receives data from node $j$; and $a_{ij}(t)=0$ if there is no data
transmission between nodes $i$ and $j$. It is natural to set
$a_{ii}(t)=0$. Clearly, the matrix $A(t)$ is antisymmetrical:
 $$a_{ij}(t)=-a_{ji}(t),\;\; t\in {\mathbf Z}^+. $$

\item[{(H$_2$)}] The network initially has two nodes at $t=0$, i.e., $n_0=2$, and
the number of nodes in the network increases by one at a time;
that is, $n_t:=t+2$ for $t>0$. The new node does not send or
receive any data from all the old nodes at time $t$; that is,
 $$a_{t+3, j}(t)=0,\;\;1\le j\le t+2,\;t\in {\mathbf Z}^+. $$

\item[{(H$_3$)}] The matrix $A(t)$ is in the following form:
 $$A(0)=\left(\begin{array}{c} J\\ 0
 \end{array} \right),\;\;A(2t)=\left(\begin{array}{cc} J&0\\ 0&0
 \end{array} \right),\;t\ge 1,                         \eqno{[4]}$$
 $$A(2t+1)=\left(\begin{array}{cc} -J & B(2t+1)\\
 -B^T(2t+1) & C(2t+1)\\ 0 & 0
 \end{array} \right),\;t\in {\mathbf Z}^+,            \eqno{[5]}$$
where
 $$B(2t+1)=\left(\begin{array}{cccc}
 b_{11} & b_{12} & \cdots & b_{1, 2t+1}\\
 b_{21} & b_{22} & \cdots & b_{2, 2t+1}
 \end{array}\right)(2t+1)$$
is a $2\times(2t+1)$ T-matrix, $C(2t+1)$ is a $(2t+1)\times(2t+1)$
antisymmetrical T-matrix, with zero block-matrices in compatible
dimensions, and
 $$J:=\left(\begin{array}{cc}
 0 & -1 \\1 & 0
 \end{array} \right). $$
\end{itemize}

It is clear that $J$ is a $2\times 2$ antisymmetrical matrix and
satisfies $J^2=-I_2$, where $I_2$ is the $2\times 2$ identity
matrix.

It follows from (H$_3$) by induction that
 $$D(2t)=\left(\begin{array}{c} J\\ 0
 \end{array} \right),\;\;
 D(2t+1)=\left(\begin{array}{c} I_2 \\
 -B^T(2t+1)J\\ 0 \end{array} \right),\;t\in {\mathbf Z}^{+}.    \eqno{[6]}$$

For any initial point $x(0)=(a,b)^T\in {\mathbf R}^2$, it follows
from ${\mathbf 2}$ and ${\mathbf 6}$ that the corresponding
solution $x(t)$ can be written as
 $$x(2t+1)=\left(\begin{array}{c} -b\\ a\\0
 \end{array} \right);\;\;
 x(2t+2)=\left(\begin{array}{c} a \\ b\\u(2t+2)\\ 0
 \end{array} \right),\;t\in {\mathbf Z}^{+},                   \eqno{[7]}$$
where
 $$\ba{ll}&u(2t+2)=-B^T(2t+1)Jx(0)\\[2.0ex]
 =&(b_{11}b-b_{21}a, b_{12}b-b_{22}a, \cdots,
 b_{1, 2t+1}b-b_{2,2t+1}a)^T(2t+1).\ea$$
Since $b_{ij}(2t+1)\in \{-1,0,1\}$, it follows from ${\mathbf 7}$
that for all $t\in {\mathbf Z}^{+}$ and for all $1\le j\le n_t$,
 $$|x_j(t)|\le |a|+|b|.                                 \eqno{[8]}$$
In addition, it follows from ${\mathbf 7}$ that
 $$\ba{ll}\Delta(x)(2t+1)=-b+a,\\[2.0ex]
 \Delta (x)(2t+2)=a \Bigl(1-\sum_{i=1}^{2t+1}b_{2i}(2t+1)\Bigr)
 +b \Big(1+\sum_{i=1}^{2t+1}b_{1i}(2t+1)\Big),\ea       \eqno{[9]}$$
which imply that
 $$\ba{ll}|\Delta(x)(2t+1)|\le |b|+|a|,\\[2.0ex]
 |\Delta (x)(2t+2)|\le
 |a|\Big|1-\sum_{i=1}^{2t+1}b_{2i}(2t+1)\Big|+
 |b|\Big|1+\sum_{i=1}^{2t+1}b_{1i}(2t+1)\Big|, \;t\in {\mathbf Z}^{+}.
 \ea                                                    \eqno{[10]}$$

As noted in Remark 1, for this computer network model, one may
choose the constant $\alpha =1/2$ in Definition 1. The discussion
on the stability of system ${\mathbf 1}$ with this choice is
divided into the following two cases: \medskip

\noindent{\bf Case I.} Suppose that the two sequences
$\Big\{\sum_{i=1}^{2t+1}b_{1i}(2t+1)\Big\}_{t=0}^{\infty}$ and
$\Big\{\sum_{i=1}^{2t+1}b_{2i}(2t+1)\Big\}_{t=0}^{\infty}$ are
bounded. Then, for any $(2t+1)\times (2t+1)$ antisymmetrical
T-matrix $C(t)$, the network is stable.

In fact, by the assumption there exists a positive constant
$\gamma$ such that
 $$\Bigg|1-\sum_{i=1}^{2t+1}b_{2i}(2t+1)\Bigg|,
 \Bigg|1+\sum_{i=1}^{2t+1}b_{1i}(2t+1)\Bigg|\le
 \gamma,\;\;t\in {\mathbf Z}^{+}.$$
Denote
 $$C_0:=\inf\{S_j:\,1\le j \le n_t,\;t\in {\mathbf Z}^{+}\}  \eqno{[11]}$$
and only consider the situation where $C_0>0$ in the following.

For any initial value $x(0)=(a,b)^T$ with
 $$|a|+|b|\le \min\{C_0, M_0, M_0/\gamma\}, $$
it follows from ${\mathbf 8}$ and ${\mathbf 10}$ that the solution
$x(t)$ satisfies
 $$\ba{ll}|x_j(t)|\le C_0\le S_j,\;\;1\le j\le n_t,\;\;
 |\Delta(x)(t)|\le M_0,\;\; t\in {\mathbf Z}^{+}.
 \ea                                                    \eqno{[12]}$$
Hence, the network is stable.
\medskip

\noindent{\bf Case II.} For any $(2t+1)\times (2t+1)$ antisymmetric
T-matrix $C(t)$, the model is a devil network if and only if at
least one of the two sequences
$\Big\{\sum_{i=1}^{2t+1}b_{1i}(2t+1)\Big\}_{t=0}^{\infty}$ and
$\Big\{\sum_{i=1}^{2t+1}b_{2i}(2t+1)\Big\}_{t=0}^{\infty}$ is
unbounded.

The necessity follows from the conclusion of Case I.

To show the sufficiency, without loss of generality, suppose that
$\Big\{\sum_{i=1}^{2t+1}b_{1i}(2t+1)\Big\}_{t=0}^{\infty}$ is
unbounded. Then, for any $a\in {\mathbf R}$ with $0<|a|\le
\min\{C_0, M_0/2\}$, there exist infinitely many $t_k\ge 1, k\ge 1$,
such that
 $$\Bigg|\,1-\sum_{i=1}^{2t_k+1}b_{2i}(2t_k+1)\,\Bigg|>M_0/|a|. $$
Consequently, by ${\mathbf 8}$ and ${\mathbf 9}$, the solution
$x(t)$ of system ${\mathbf 1}$ with the initial value
$x(0)=(0,a)^T$ satisfies that, for all $t\ge 0$,
 $$\ba{ll}|x_j(t)|=|a|\le C_0\le S_j,\;\;1\le j\le n_t,\\[2.0ex]
 |\Delta(x)(2t+1)|=|a|\le M_0/2, \ea $$
and
 $$|\Delta (x)(2t_k+2)|=|a|
 \Bigg|\,1-\sum_{i=1}^{2t_k+1}b_{2i}(2t_k+1)\,
 \Bigg|>M_0. $$
Therefore, the model is a devil network.

To this end, it should be noted that in the second case above, for
some initial values the storages of the server (or the whole
network) can oscillate more and more strongly as time evolves.

\bigskip

\noindent {\bf Example.} Consider the special case of
$B(2t+1)=(B_1(2t+1),0)$, where $0$ is a $2\times t$ zero matrix and
 $$B_1(2t+1)= \left(\begin{array}{cccc}
 0 & 0 & \cdots & 0 \\
 -1 & -1 & \cdots & -1 \end{array}\right)$$
is a $2\times (t+1)$ T-matrix. It is clear that $B(2t+1)$
satisfies the condition given in Case II above. So, the network is
a devil network. It follows from ${\mathbf 7}$ that
 $$x(2t+1)=(b, -a, 0)^T,\;\;
 x(2t+2)=(a,b,a,\cdots, a, 0)^T,\;t\in {\mathbf Z}^{+}, $$
where $0$ in the first relation is a $(2t+2)$-dimensional zero row
vector and $0$ in the second relation is a $(t+1)$-dimensional
zero row vector. Hence, for all $t\in {\mathbf Z}^{+}$, one has
$|x_j(t)|\le \max\{|a|, |b|\}$, $1\le j\le n_t$, and
 $$|\Delta (x)(2t+1)|= |b-a|,\;\; |\Delta (x)(2t+2)|=|b+(t+2)a|.
                                                    \eqno{[13]}$$

Obviously, in the case of $a\ne 0$, the storage of the server
strongly oscillates as time evolves. This illustrates that the
network runs quite well at some times, but will break down at some
other times, when time is sufficiently large.

Moreover, the matrix $B(2t+1)$ in this example describes the
phenomenon that the second computer in the network continuously
sends data $a\ne 0$ to each of the other $t+1$ computers. Although
the burden of each of these $t+1$ computers received from this
second computer is equal to $a$, which is not heavy if $|a|$ is
small, it gives an extra load to the server. If the network is an
email system, this example explains why a virus-contaminated
computer can cause the server to breakdown since burden is
continuously building up on the server in this way as described by
the new model.
\medskip

\noindent {\bf Remark 2.}\begin{itemize}
\item[{(1)}]
Since the coupling matrix $A(t)$ of the computer network discussed
in this section is a T-matrix, all the entries of $D(t)$ defined
by ${\mathbf 6}$ are integers. So, system ${\mathbf 1}$ cannot be
chaotic in the sense of Li-Yorke. However, when the connections
are weighted or time-varying, the linear system ${\mathbf 1}$ may
become chaotic in the sense of Li-Yorke (5), which will be further
discussed elsewhere in the near future.
\item[{(2)}] In the above example, if $y=\Delta (x)$ is taken as an
output of the system, the output according to ${\mathbf 13}$ is
chaotic in the sense of Li-Yorke. In fact, in this case, there is
an uncountable scrambled set in the diagonal line $\{(a,a):\;a\in
{\mathbf R}\}$ (6).
\end{itemize}

\bigskip

\noindent {\large \bf 3.\, Stability for a general linear model of
networks} \vspace{0.10in}

Consider the stability of a general model of networks, i.e., its
corresponding system ${\mathbf 1}$ is linear, in which its
connections may be directed and weighted, and its dimension,
connectivity as well as weights may vary with time.

Let $n_t$ be the number of all the nodes in the network at time
$t$. Suppose that $x_j(t)$ represents a quantity of some property
$\cal P$ of node $j$ at time $t$, $1\le j\le n_t$, and
$x(t)=(x_1(t),x_2(t),\ldots, x_{n_t}(t))^T\in {\mathbf R}^{n_t}$
satisfies the linear system ${\mathbf 1}$, where
$A(t)=(a_{ij}(t))$ is an $n_{t+1}\times n_t$ matrix and its entry
$a_{ij}(t)$ represents a weight, which is no longer restricted to
the set of $\{-1,0,1\}$, with direction from node $i$ to node $j$
at time $t$.

Similarly assume that each node $i$ in the network has its own
maximum quantity (e.g., storage) $S_i$, invariant in time, and the
whole network has its own maximum quantity which may be infinite or
varying with time, for the property $\cal P$.

The following discussion is divided into two cases: (1) the maximum
quantity for property $\cal P$ of the whole network is
time-invariant, which can be either finite or infinite; (2) the
maximum quantity for property $\cal P$ of the whole network is
time-varying.

\bigskip \newpage

\noindent {\bf 3.1.\, Networks with time-invariant maximum
quantity of property $\cal P$} \vspace{0.10in}

Let $M_0$ be the maximum quantity for property $\cal P$ of the whole
network, which is a positive constant or infinity. In this case, the
definitions of stable, unstable, and devil networks are similar to
those given in Definition 1 in Section 2.

It is clear that for any given initial point $x(0)\in {\mathbf
R}^{n_0}$, the solution $x(t)$ of system ${\mathbf 1}$ can also be
written as ${\mathbf 2}$, with $D(t)=(d_{ij}(t))_{n_{t+1}\times
n_0}$ satisfying ${\mathbf 3}$.

Next, the stability and instability of system ${\mathbf 1}$ are
studied for the case of $M_0<\infty$. \medskip

\noindent {\bf Theorem 1.} Assume that the maximum quantities
$M_0$ and $S_i$ for property $\cal P$ of the whole network and of
each node $i$ are both finite. Then, the network described by
system ${\mathbf 1}$ is stable if and only if
$\Big\{\sum_{i=1}^{n_{t+1}}d_{ij}(t)\Big\}_{t=0}^{\infty}$ is
bounded for all $1\le j\le n_0$, and moreover there exists a
positive constant $\beta$ such that
$$|d_{ij}(t)|\le \beta S_i,\;\; 1\le j\le n_0,\;1\le i\le n_{t+1},\;
t\in {\mathbf Z}^{+}.                          \eqno{[14]}$$

\medskip

\noindent {\bf Proof.} First, the sufficiency is verified. By the
assumption, there exists a constant $\gamma>0$ such that
$$\Bigg|\,\sum_{i=1}^{n_{t+1}}d_{ij}(t)\,\Bigg| \le \gamma,\;
 1\le j\le n_0, \;t\in {\mathbf Z}^{+}.              \eqno{[15]}$$
It follows from ${\mathbf 2}$, ${\mathbf 14}$, and ${\mathbf 15}$
that, for all $t\in {\mathbf Z}^{+}$,
 $$|x_i(t+1)|
 \le \sum_{j=1}^{n_0}|d_{ij}(t)||x_j(0)| \le \beta S_i
 \sum_{j=1}^{n_0}|x_j(0)|,\;1\le i\le n_{t+1}, $$
and
 $$\ba{ll} &|\Delta(x)(t+1)|
 = \Big|\sum_{i=1}^{n_{t+1}}\Big(\sum_{j=1}^{n_0}d_{ij}(t)x_j(0)
 \Big)\Big|\\[2.0ex]
 \le&
 \sum_{j=1}^{n_0}\Big|\sum_{i=1}^{n_{t+1}}d_{ij}(t)\Big||x_j(0)|\le
 \gamma \sum_{j=1}^{n_0}|x_j(0)|.\ea               \eqno{[16]} $$
So, for any initial point $x(0)\in {\mathbf R}^{n_0}$ with
$\sum_{j=1}^{n_0}|x_j(0)|\le r_0$, where
$$r_0=\min\{S_1, S_2,\ldots, S_{n_0}, 1/\beta, M_0, M_0/\gamma\},
                                                    \eqno{[17]}$$
one has
$$|x_i(t)|\le S_i,\;\;1\le i\le n_t,\;\;
 |\Delta(x)(t)|\le M_0,\;t\in {\mathbf Z}^{+}.    \eqno{[18]} $$
Hence, the network is stable.

Then, the necessity is verified. Since the network is stable,
there exists a positive constant $r_0$ such that for any initial
point $x(0)\in {\mathbf R}^{n_0}$ with
$\sum_{j=1}^{n_0}|x_j(0)|\le r_0$ one has $|x_i(t+1)|\le
S_i,\;1\le i\le n_{t+1}$, and $|\Delta(x)(t+1)|\le M_0$, for all
$t\in {\mathbf Z}^{+}$. Given any $j_0$, $1\le j_0\le n_0$, set
$x_{j_0}(0)=r_0$ and $x_j(0)=0,\;1\le j\ne j_0\le n_0$. Then, it
follows from ${\mathbf 2}$ that, for all $t\in {\mathbf Z}^{+}$,
 $$\ba{ll}|x_i(t+1)|=|d_{ij_0}(t)x_{j_0}(0)|=r_0\,|d_{ij_0}(t)|
 \le S_i,\;\;1\le i\le n_{t+1},\\[2.0ex]
 |\Delta(x)(t+1)|=\Big|\sum_{i=1}^{n_{t+1}}d_{ij_0}(t)x_{j_0}(0)
 \Big|=r_0\,\Big|\sum_{i=1}^{n_{t+1}}d_{ij_0}(t)\Big|\le M_0,\ea$$
which implies that, for all $t\in {\mathbf Z}^{+}$,
 $$|d_{ij_0}(t)|\le S_i/r_0, \;1\le i\le n_{t+1},\; \;
 \Bigg|\,\sum_{i=1}^{n_{t+1}}d_{ij_0}(t)\,\Bigg|\le M_0/r_0. $$
Hence, inequality ${\mathbf 14}$ holds with $\beta=1/r_0$, and
$\Big\{\sum_{i=1}^{n_{t+1}}d_{ij}(t)\Big\}_{t=0}^{\infty}$ is
bounded for all $1\le j\le n_0$. The necessity is thus verified.

Therefore, the proof is complete.

\medskip

\noindent {\bf Theorem 2.} Assume that the maximum quantities
$M_0$ and $S_i$ for property $\cal P$ of a whole network and each
node $i$ are both finite. Then, the network described by system
${\mathbf 1}$ is a devil network if there exists a positive
constant $\beta$ such that
$$|d_{ij}(t)|\le \beta S_i,\;\; 1\le j\le n_0, \;1\le i\le
n_{t+1},\;t\in {\mathbf Z}^{+},                               $$
and moreover there exist two time subsequences,
$\{t_k\}_{k=1}^{\infty}$ and $\{t_k'\}_{k=1}^{\infty}$ with
$t_k\to \infty$ and $t_k'\to \infty$ as $k\to \infty$, such that
$\Big\{\sum_{i=1}^{n_{t_k+1}}d_{ij}(t_k)\Big\}_{k=1}^{\infty}$ is
bounded for all $1\le j\le n_0$ and
$\Big\{\sum_{i=1}^{n_{t_k'+1}}d_{ij_0}(t_k')\Big\}_{k=1}^{\infty}$
is unbounded for some $1\le j_0\le n_0$.

\medskip

\noindent {\bf Proof.} Since
$\Big\{\sum_{i=1}^{n_{t_k+1}}d_{ij}(t_k)\Big\}_{k=1}^{\infty}$ is
bounded for all $1\le j\le n_0$, there exists a constant $\gamma
>0$ such that
 $$\Bigg|\,\sum_{i=1}^{n_{t_k+1}}d_{ij}(t_k)\,\Bigg|\le \gamma,
 \;\;k\ge 1,\; 1\le j\le n_0.$$
By assumption,
$\Big\{\sum_{i=1}^{n_{t_k'+1}}d_{ij_0}(t_k')\Big\}_{k=1}^{\infty}$
is unbounded for some $1\le j_0\le n_0$. Without loss of generality,
suppose that
 $$\Bigg|\,\sum_{i=1}^{n_{t_k'+1}}d_{ij_0}(t_k')\,\Bigg|\to \infty \;\;
 {\rm as}\;k\to \infty.$$
Then, for any positive constant $r\le\min\{S_1, S_2,\ldots,
S_{n_0}, 1/\beta, M_0, \alpha M_0/\gamma\}$, where
$\alpha,\,0<\alpha <1$, is the scaling parameter for system
${\mathbf 1}$, there exists $k_0\ge 1$ such that
 $$\Bigg|\,\sum_{i=1}^{n_{t_k'+1}}d_{ij_0}(t_k')\,\Bigg|> M_0/r,
 \;\;k\ge k_0.$$
Choose an initial point $x(0)\in {\mathbf R}^{n_0}$ with
$x_{j_0}(0)=r$ and $x_j(0)=0$ for all $1\le j\ne j_0\le n_0$. It is
clear that $\sum_{j=1}^{n_0}|x_j(0)|=r$. With an argument similar to
that used in the proof of the sufficiency of Theorem 1, one can
easily show that the corresponding solution $x(t)$ satisfies
 $$\ba{ll}|x_i(t)|\le S_i,\;\;1\le i\le n_t,\;\;t\in {\mathbf Z}^{+},\\[2.0ex]
 |\Delta(x)(t_k+1)|\le \alpha\,M_0,\;\;k\ge 1,\;\; |\Delta(x)(t_k'+1)|>
 M_0,\;\;k\ge k_0. \ea $$
Therefore, the network described by system ${\mathbf 1}$ is a
devil network. This completes the proof.

\medskip

In the case of $M_0=\infty$, the following result can be easily
verified by an argument similar to that used in the proof of
Theorem 1.
\medskip

\noindent {\bf Theorem 3.} Assume that the maximum quantity $S_i$
of property $\cal P$ for each node $i$ in a network is finite and
the maximum quantity for property $\cal P$ of the whole network is
infinite. Then, the network described by system ${\mathbf 1}$ is
stable if and only if there exists a constant $\beta>0$ such that
$$|d_{ij}(t)|\le \beta S_i,\;\; 1\le j\le n_0,\;
 1\le i\le n_{t+1},\;t\in {\mathbf Z}^{+}. $$

\medskip

The model discussed in Section 2 is revisited here based on the
results obtained above. In this model, $n_0=2$, and it follows
from ${\mathbf 6}$ that, for all $t\in {\mathbf Z}^{+}$,
 $$\ba{ll}|d_{ij}(t)|\le 1\le S_i/C_0,\;\; 1\le i\le t+3,\;\; j=1,2,\\[2.0ex]
 \sum_{i=1}^{2t+3}d_{i1}(2t)=1,\;\;\sum_{i=1}^{2t+3}d_{i2}(2t)=-1,\\[2.0ex]
 \sum_{i=1}^{2t+4}d_{i1}(2t+1)=1-\sum_{i=1}^{2t+1}b_{2i}(2t+1),\;\;
 \sum_{i=1}^{2t+4}d_{i2}(2t+1)=1+\sum_{i=1}^{2t+1}b_{1i}(2t+1),\ea$$
where $C_0$ is defined by ${\mathbf 11}$. Therefore, by Theorem 1,
this network is stable if and only if the two sequences
$\Big\{\sum_{i=1}^{2t+1}b_{1i} (2t+1)\Big\}_{t=0}^{\infty}$ and
$\{\sum_{i=1}^{2t+1}b_{2i}(2t+1)\}_{t=0}^{\infty}$ are bounded.
Further, by Theorem 2, this network is a devil network if and only
if at least one of the two sequences,
$\Big\{\sum_{i=1}^{2t+1}b_{1i}(2t+1)\Big\}_{t=0}^{\infty}$ and
$\Big\{\sum_{i=1}^{2t+1}b_{2i}(2t+1)\Big\}_{t=0}^{\infty}$, is
unbounded. These conclusions are the same as those obtained in
Section 2.

\bigskip

\noindent {\bf 3.2.\, Networks with time-varying maximum quantity
for property $\cal P$} \vspace{0.10in}

Let $M(t)$ be a maximum quantity of property $\cal P$ for the whole
network at time $t$.

In this case, the definitions of stable and unstable networks are
also similar to those given in Definition 1. For convenience, they
are rephrased as follows.
\medskip

\noindent {\bf Definition 2.} A network is said to be stable if
there exists a positive constant $r_0$ such that for all initial
points $x(0)=(x_1(0),x_2(0),\ldots,x_{n_0})^T\in {\mathbf
R}^{n_0}$ with $\sum_{i=1}^{n_0} |x_i(0)|\le r_0$, one has
$|x_i(t)|\le S_i, \;1\le i\le n_t$, and $|\Delta(x)(t)|\le M(t)$
for all $t\in {\mathbf Z}^+$. Otherwise, it is said to be
unstable. In particular, a network is called a devil network if it
is unstable and, moreover, for any small positive constant $r$
there exists an initial point $x(0)\in {\mathbf R}^{n_0}$ with
$\sum_{i=1}^{n_0}|x_i(0)|< r$ such that $|x_i(t)|\le S_i, \;1\le
i\le n_t$, for all $t\in {\mathbf Z}^+$, $|\Delta(x)(t_k)|\le
\alpha M(t_k)$ for infinitely many times $t_k$, for some constant
$0<\alpha < 1$, and $|\Delta(x)(t_k')|> M(t_k')$ for infinitely
many times $t_k'$, $k\ge 1$.
\medskip

\noindent {\bf Theorem 4.} Assume that the maximum quantity $S_i$
of property $\cal P$ for each node $i$ in the network is finite,
and the maximum quantity $M(t)$ of property $\cal P$ for the whole
network is finite, at any time $t\in {\mathbf Z}^{+}$. Then, the
network described by system ${\mathbf 1}$ is stable if and only if
there exist positive constants $\beta$ and $\gamma$ such that
 $$|d_{ij}(t)|\le \beta S_i,\;\; 1\le i\le n_{t+1},\;\;
 \Bigg|\,\sum_{i=1}^{n_{t+1}}d_{ij}(t)\,\Bigg|\le \gamma M(t+1),\;
 1\le j\le n_0,\;t\in {\mathbf Z}^{+}. $$

\medskip

\noindent {\bf Proof.} The proof of the theorem is similar to that
of Theorem 1, except the following: (i) in the proof of the
sufficiency, $\gamma$ in ${\mathbf 15}$ and ${\mathbf 16}$ is
replaced by $\gamma M(t+1)$, $r_0$ in ${\mathbf 17}$ is replaced
by
 $$r_0=\min\{S_1,S_2,\ldots,S_{n_0},M(0),1/\beta,1/\gamma\},$$
and $M_0$ in ${\mathbf 18}$ is replaced by $M(t)$; (ii) in the
proof of the necessity, $M_0$ is replaced by $M(t+1)$.

This completes the proof.

\medskip

Similar to Theorem 2, the following result can be established.
\medskip

\noindent {\bf Theorem 5.} Assume that the maximum quantity $S_i$
of property $\cal P$ for each node $i$ in the network is finite,
and a maximum quantity $M(t)$ of property $\cal P$ for the whole
network is finite, at any time $t\in {\mathbf Z}^{+}$. Then, the
network described by system ${\mathbf 1}$ is a devil network if
there exists a positive constant $\beta$ such that
$$|d_{ij}(t)|\le \beta S_i,\;\; 1\le j\le n_0,\;1\le i\le n_{t+1},\;
 t\in {\mathbf Z}^{+},                                      $$
and moreover there exist two time subsequences,
$\{t_k\}_{k=1}^{\infty}$ and $\{t_k'\}_{k=1}^{\infty}$ with
$t_k\to \infty$ and $t_k'\to \infty$ as $k\to \infty$, such that
$\Big\{M^{-1}(t_k+1)\sum_{i=1}^{n_{t_k+1}}d_{ij}(t_k)\Big\}_{k=1}^{\infty}$
is bounded for all $1\le j\le n_0$ and
$\Big\{M^{-1}(t_k'+1)\sum_{i=1}^{n_{t_k'+1}}d_{ij_0}(t_k')\Big\}_{k=1}^{\infty}$
is unbounded for some $1\le j_0\le n_0$.

\medskip

\noindent {\bf Proof.} The proof of the theorem is similar to that
of Theorem 2. So the details are omitted.

\medskip

\noindent {\bf Remark 3.}
\begin{itemize}
\item[(1)] If the number of nodes in a network does not vary; that is,
the dimension $n_t$ of the corresponding system ${\mathbf 1}$ is
time-invariant, then system ${\mathbf 1}$ is a classical
time-varying discrete linear system, which surprisingly can be
chaotic in the sense of Li-Yorke (5).

\item[(2)] If the effects of internal and external nonlinearities
on a network are considered, the corresponding system is by nature
nonlinear, for which the stability and complex dynamical behaviors
need to be further addressed in the future.
\end{itemize}

\noindent {\bf Remark 4.} In the present letter, only the simplest
possible model of a localized isolated computer network with one
server is considered. A realistic computer model, however, has
more than one server in general, which becomes more mathematically
involved, leaving a challenging topic for future research.

\vspace{0.3in}

\noindent {\large \bf References}

\begin{enumerate}
\vspace{-0.05in}
\item[{1.}] Watts DJ, Strogatz SH (1998) {\it Nature}
393: 440--442.

\vspace{-0.05in}
\item[{2.}] Barab\`asi AL, Albert R (1999) {\it Science}
286: 509--512.

\vspace{-0.05in}
\item[{3.}] Erdos P, Renyi A (1960) {\it
Publ. Math. Inst. Hung. Acad. Sci.} 5: 17--60.

\vspace{-0.05in}
\item[{4.}] Robinson C (1995) {\it Dynamical systems: stability,
symbolic dynamics and chaos} (CRC Press, Florida).

\vspace{-0.05in}
\item[{5.}] Shi Y, Chen G (2006) {\it Chaos of time-varying discrete
dynamical systems}, under review.

\vspace{-0.05in}
\item[{6.}] Shi Y,  Yu P (2006) {\it Chaos, Solitons and Fractals}
28: 1165--1180.

\end{enumerate}

\end{document}